\documentclass[
reprint,
amsmath,
amssymb,
aip,
pop,
]{revtex4-1}

\usepackage{graphicx}
\usepackage{upgreek}
\usepackage{color}
\usepackage[normalem]{ulem}

\renewcommand{\mu}{\upmu}

\begin{document}

\title{Accurate modeling of plasma acceleration with arbitrary order pseudo-spectral
 particle-in-cell methods}
 \label{numerical-growth-of-energy-spread-in-pwfa-particle-in-cell-simulations}
\author{S. Jalas}
\email{soeren.jalas@desy.de}
\affiliation{Center for Free-Electron Laser Science \& Department of Physics, 
  University of Hamburg, 22761 Hamburg, Germany}
\author{I. Dornmair}
\affiliation{Center for Free-Electron Laser Science \& Department of Physics,
  University of Hamburg, 22761 Hamburg, Germany}
\author{R. Lehe}
\affiliation{Lawrence Berkeley National Laboratory, Berkeley, CA 94720, USA}
\author{H. Vincenti}
\affiliation{Lawrence Berkeley National Laboratory, Berkeley, CA 94720, USA}
\author{J.-L. Vay}
\affiliation{Lawrence Berkeley National Laboratory, Berkeley, CA 94720, USA}
\author{M. Kirchen}
\affiliation{Center for Free-Electron Laser Science \& Department of Physics,
  University of Hamburg, 22761 Hamburg, Germany}
\author{A. R. Maier}
\affiliation{Center for Free-Electron Laser Science \& Department of Physics,
  University of Hamburg, 22761 Hamburg, Germany}

\begin{abstract}
Particle in Cell (PIC) simulations are a widely used tool for the investigation
 of both laser- and beam-driven plasma acceleration. It is a known issue that 
 the beam quality can be artificially degraded by numerical Cherenkov radiation
 (NCR) resulting primarily from an incorrectly modeled dispersion relation. Pseudo-spectral 
 solvers featuring infinite order stencils can strongly reduce NCR -- or
 even suppress it -- and are 
therefore well suited to correctly model the beam properties. For efficient 
 parallelization of the PIC algorithm, however, localized solvers are 
 inevitable. Arbitrary order pseudo-spectral methods provide this needed locality. 
 Yet, these methods can again be prone to NCR. Here, we show that acceptably 
 low solver orders are sufficient to correctly model the physics of interest, 
 while allowing for parallel computation by domain decomposition.

\end{abstract}

\maketitle

\section*{Introduction}\label{introduction}
Plasma accelerators provide high accelerating gradients and promise very
compact next-generation particle accelerators as drivers for brilliant 
light sources \cite{Maier,Huang} and high energy physics \cite{Schroeder}. 
Using laser or beam drivers, the acceleration of electron beams by multiple 
GeV over few-cm distances has been demonstrated in experiments 
\cite{Leemans,PhysRevLett.113.245002,Litos,Texas}. The improvement of beam quality, especially in 
terms of energy spread, remains crucial for applications. Due to the non-linear 
nature of the beam-plasma interaction, Particle-in-Cell (PIC) simulations are 
a vital tool to model the beam injection and transport in the plasma. PIC 
algorithms self-consistently solve the interaction of spatially discretized 
electromagnetic fields with charged particles defined in a continuous phase 
space.
However, these codes can suffer from numerical Cherenkov radiation (NCR) 
\cite{GODFREY} that is caused primarily by incorrect modeling of the electromagnetic 
dispersion relation. Mitigation of this effect is crucial as 
it gives rise to an unphysical degradation of beam quality which leads to wrong
predictions when using those results as input for studies on applications such as 
compact Free-Electron Lasers. For example, NCR has been found to artificially decrease 
the beam quality in terms of transverse beam emittance \cite{lehe2013} for 
finite-difference time domain (FDTD) solvers, such as the commonly used Yee 
scheme. Efforts to circumvent this issue include the modification of the FDTD 
dispersion relation \cite{lehe2013,PhysRevSTAB.16.041303} and digital filtering 
\cite{Greenwood,Vay20115908}. 
In contrast to FDTD solvers, pseudo-spectral solvers, which solve Maxwell's 
equations in spectral space, strongly reduce NCR -- and sometimes even
suppress it.\\
In practice, PIC simulation can easily demand thousands of hours of computation 
time. Therefore, implementations that are efficient and scalable for parallel 
production are unavoidable. The parallelization of PIC codes is typically done 
by decomposing the simulation volume into domains which are evaluated by 
individual processes.
For FDTD algorithms efficient scaling to several thousand parallel processes is 
possible. Pseudo-spectral solvers, however, act globally on the grid due to the 
needed Fourier transform, which limits efficient parallelization. An approach 
to overcome this limitation are \textit{arbitrary order} pseudo-spectral solvers 
\cite{VayPSAOTD,Vincenti,li2016controlling}, 
which only locally affect the fields when solving Maxwell's equations.
On the downside, this --- again --- introduces spurious numerical dispersion and can 
make a simulation prone to NCR.\\
In the following, we investigate numerical Cherenkov radiation and its 
implications on the physics of a simulation for arbitrary order pseudo-spectral 
solvers using the spectral, quasi-cylindrical PIC algorithm implemented in the 
code \textsc{Fbpic} \cite{Lehe201666,fbpic_repo}.\\
The paper is structured as follows. We will first summarize the concept of 
pseudo-spectral solvers, with an emphasis on the arbitrary order \cite{VayPSAOTD}
\textit{pseudo-spectral analytic time domain} (PSATD) \cite{Haber}, 
followed by a discussion of the main characteristics of NCR. In the last part 
we numerically analyze the mitigation of NCR in the arbitrary order PSATD 
scheme. As a figure of merit for the physical accuracy of the solver we use the 
effect of NCR on the beam emittance and the slice energy spread. With parameter
scans we show the dependence of NCR on physical and numerical parameters of the 
simulation.

\section*{Finite order pseudo-spectral solvers}\label{psatd-finite-order-stencil-in-fbpic}
A widely used method in PIC codes are FDTD solvers. These algorithms use finite 
difference operators to approximate derivatives in time and space when numerically 
solving Maxwell's equations, which introduces a spurious numerical 
dispersion. This approximation can be done with arbitrary precision by 
increasing the used stencil order, i.e., taking contributions from more 
surrounding grid nodes into account. Yet, this has the drawback of also 
increasing the computational demand of the calculation. Further, common FDTD 
algorithms represent the electromagnetic fields on a staggered grid, where the 
electric and magnetic fields are not defined at the same points in time and 
space. This is known to cause errors when modeling the direct interaction 
between an electron beam and a co-propagating laser \cite{lehe2014}.\\
The use of finite differences in space can be avoided by transforming Maxwell's 
equations into spectral space. In that case the differential operator 
$\partial_z$ is given by a multiplication with $ik_z$, which numerically can be 
evaluated without approximations. The resulting solver class is referred to as 
\textit{pseudo-spectral time domain} (PSTD) \cite{pstd} and it describes the 
limit of an FDTD solver with an infinite stencil order \cite{Vincenti}. Yet, 
this solver still applies finite difference approximations in time and thereby, 
like FDTD solvers, its timestep is limited by a Courant condition. It is 
subject to spurious numerical dispersion, but unlike FDTD schemes the erroneous 
dispersion caused by the PSTD scheme results in modes traveling faster than the 
speed of light \cite{Vay2013260}.\\
Under the assumption of constant currents $\boldsymbol{J}$ over one timestep $\Delta t$, this 
constraint can be overcome by integrating Maxwell's equations analytically in 
time. With a precise representation of derivatives in time and 
space, the resulting \textit{pseudo-spectral analytic time domain} (PSATD) 
 scheme is not subject to a Courant condition and is free of 
spurious numerical dispersion. Furthermore, it is currently the only solver that can be 
combined with the Galilean scheme \cite{kirchen2016,lehe_galilean} that is
intrinsically stable when modeling relativisticly drifting plasmas with uniform velocity.\\ 
The PSATD field propagation equations in Cartesian coordinates are 

\begin{subequations}
\begin{align}
    \boldsymbol{\tilde{E}}^{n+1}&=C\boldsymbol{\tilde{E}}^{n}
    +iS\boldsymbol{\hat{k}}\times\boldsymbol{\tilde{B}}^{n}
    -\frac{S}{ck}\boldsymbol{\tilde{J}}^{n+1/2} \label{psatd-eqE}\\ \nonumber
    &+i\frac{\boldsymbol{\hat{k}}}{k}\left[\left(\frac{S}{ck\Delta t}-1\right)\tilde{\rho}^{n+1}+\left(C-\frac{S}{ck\Delta t} \right) \tilde{\rho}^n  \right]  
    \\
    \boldsymbol{\tilde{B}}^{n+1}&=C\boldsymbol{\tilde{B}}^{n}
    -iS\boldsymbol{\hat{k}}\times\boldsymbol{\tilde{E}}^{n}
    +i\frac{1-C}{ck}\boldsymbol{\hat{k}}\times\boldsymbol{\tilde{J}}^{n+1/2},  \label{psatd-eqB}
\end{align}
\end{subequations}

where $\boldsymbol{k}$ is the wave vector of length $k\equiv\sqrt{k_x^2 +
  k_y^2 + k_z^2}$ with 
$\boldsymbol{\hat{k}}=\boldsymbol{k}/k$ and $\tilde{F}$ is the Fourier transform of a 
field $F$, with $E$ and $B$ the electric and magnetic field, and
$\rho$ and $J$ the charge and current density.
 Further, $C=\cos(kc\Delta t)$ and $S=\sin(kc\Delta t$). 
With pseudo-spectral methods the fields are naturally 
centered in space and time, eliminating a source of interpolation errors due to 
the usual staggering of the electromagnetic fields in the popular Yee FDTD solver. 
The equations of the PSTD scheme can be retrieved from those of the PSATD scheme, 
by writing the PSATD equations in an equivalent staggered form \cite{Vay2013260} 
and performing a first-order Taylor expansion in $\Delta t$.\\ 
In the following we will use the PSATD scheme in a quasi-cylindrical geometry 
as derived in Ref.\ \cite{Lehe201666}. In this geometry the fields are 
decomposed into azimuthal modes $m$, which can then be 
represented on separate 2D grids \cite{Lifschitz20091803}. In practice, as the contributions from most 
of the modes go to zero for near-symmetrical systems, only a few of these have 
to be taken into account. Consequently, this geometry allows to model 
near-symmetrical problems at a drastically reduced computational effort 
compared to full 3D simulations. In the quasi-cylindrical PSATD scheme the real-space 
components $F_r$, $F_\theta$ and $F_z$ of a field $\boldsymbol{F}$ are connected 
to their spectral components $\tilde{F}_{+,m}$, $\tilde{F}_{-,m}$ and $\tilde{F}_z$ 
by the combination of a Hankel transform (along $r$) and a Fourier transform 
(along $z$). The equations of the quasi-cylindrical PSATD can be derived from 
the Cartesian PSATD scheme by replacing the differential operators 
in Eqs.\ \eqref{psatd-eqE} and \eqref{psatd-eqB} by \cite{lehe_galilean}

\begin{subequations}
\begin{align}
\boldsymbol{\tilde{F}}=i\boldsymbol{k}\tilde{S} &\to 
  \begin{pmatrix}
    \tilde{F}_{+,m}\\
    \tilde{F}_{-,m}\\
    \tilde{F}_{z,m}
  \end{pmatrix}
  =
  \begin{pmatrix}
     -k_\perp\tilde{S}_m/2\\
     k_\perp\tilde{S}_m/2\\
     ik_z\tilde{S}_m
  \end{pmatrix}
\\
\boldsymbol{\tilde{F}}=i\boldsymbol{k}\times\boldsymbol{\tilde{V}} &\to 
  \begin{pmatrix}
    \tilde{F}_{+,m}\\
    \tilde{F}_{-,m}\\
    \tilde{F}_{z,m}
  \end{pmatrix}
  =
  \begin{pmatrix}
    k_z\tilde{V}_{+,m}-ik_\perp \tilde{V}_{z,m}/2\\
    -k_z\tilde{V}_{-,m}-ik_\perp \tilde{V}_{z,m}/2\\
    ik_\perp\tilde{V}_{+,m}+ik_\perp\tilde{V}_{-,m},
  \end{pmatrix}
\end{align}
\end{subequations}

where $\tilde{S}$ is any scalar field and $\boldsymbol{\tilde{V}}$ any vector field. 
This representation of the PSATD scheme preserves the advantages of the Cartesian 
PSATD scheme, i.e. an ideal dispersion relation, a centered representation of 
the fields and no Courant condition for the timestep, and combines them with 
the reduced computational demand associated with a quasi-cylindrical geometry.\\
The propagation of the fields in spectral space in Eqs.\ \eqref{psatd-eqE} and 
\eqref{psatd-eqB}  corresponds to a global operation in spatial space, where 
information is transferred over the entire simulation grid. However, for domain 
decomposition a locality of the solver operations is necessary.
In order to achieve an arbitrary order local stencil the solver equations are 
modified.\\
The formulations of the modified pseudo-spectral scheme can be derived from 
the expression of a finite difference with arbitrary order. For a centered 
scheme the discrete derivative with stencil order $2\,n$ of a function $F$ at 
a discrete point $p$ on the spatial grid is \cite{fornberg1988generation} 

\begin{align}
    (D_z F)_p&=\sum_{j=1}^{n} 
\alpha_{n,j} \frac{F_{p+j}-F_{p-j}}{2\,j\,\Delta z} \label{eq:finite_difference}
\\ \alpha_{n,j}&=(-1)^{j+1}\frac{ 2\, (n!)^2 }{( n - j)! \;( n + j)!}. \nonumber
\end{align}
To get a spectral method with a finite stencil derivative consider the spectral 
representation of this operator
\begin{equation}
    (D_z F)_p = \sum_{k_z} i[k_z] \tilde{F}_{k_z} e^{ik_zp\Delta z}.
\end{equation}
Here $[k_z]$ is the modified wavenumber given as
\begin{equation}
  [k_z] = \sum_{j=1}^{n} 
 \alpha_{n,j} \frac{\sin(k_z j \Delta z)}{j\,\Delta z},
\end{equation}

with the grid spacing $\Delta z$ and $\tilde{F}$ the spectral form of $F$. 
This means that the equivalent to a finite stencil in real 
space can be achieved by using a modified PSTD algorithm, whereby
$k_z$, $k_x$, $k_y$ are replaced by $[k_z]$, $[k_x]$, $[k_y]$ in
the discretized Maxwell equations in spectral space. Because this modified PSTD
algorithm is mathematically equivalent to an FDTD scheme (with
high-order spatial derivatives), it has the same locality.\\
Similarly to the PSTD algorithm, in order to obtain a PSATD scheme
with improved locality, we follow identical steps. More precisely, the
discretized Maxwell equations Eqs.\ \eqref{psatd-eqE} 
and \eqref{psatd-eqB} are modified by replacing $k_z$, $k_x$, $k_y$
by $[k_z]$, $[k_x]$, $[k_y]$ -- including in the expression of $k \equiv \sqrt{k_x^2 +
  k_y^2 + k_z^2}$ which in turn impacts the expressions of the
coefficients $C$ and $S$. Consequently, unlike the modified PSTD algorithm, 
the modified PSATD scheme is not strictly local. In order to
evaluate its locality, we get the corresponding 
stencil coefficients on the spatial grid, by applying an inverse Fourier transform 
to the operators used to advance the fields in Eqs.\ \eqref{psatd-eqE} 
and \eqref{psatd-eqB}, i.e.\ $\boldsymbol{\hat{[k]}}S$ and $C$.\\
\begin{figure}[tb]
\centering
\includegraphics[width=\columnwidth]{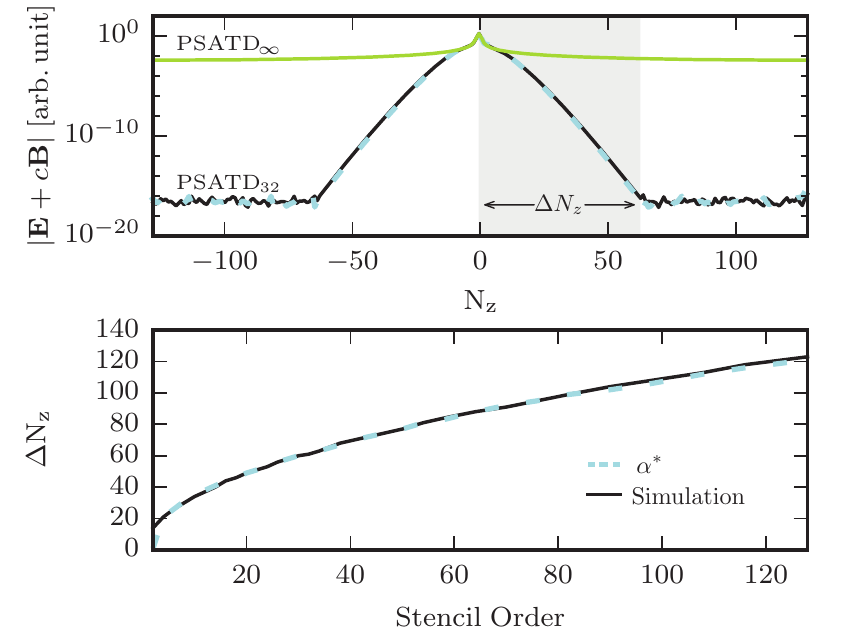}
\caption{Expansion of an electromagnetic unit pulse after one PIC cycle with 
modified $k_z$ (top). The field pushed with the unmodified PSATD method 
(green line) shows non-negligible field contributions along the entire 
simulation grid, while the fields modeled with order 32 PSATD (black line) 
decrease by $16$ orders of magnitude over $\Delta N_z$ cells and reach the 
machine precision level. The number of cells over which the fields decrease 
(bottom) depends on the stencil order. The dashed lines show the stencil 
coefficients derived from the modified PSATD equations.}
\label{fig:stencil_reach}
\end{figure}
Next, we will use this scheme in quasi-cylindrical geometry with the PIC 
framework \textsc{Fbpic} to verify the locality of the solver. In 
\textsc{Fbpic} only $k_z$ is modified, because the Fourier transform is only
used along the $z$-axis. Therefore, the 
corresponding solver stencil remains infinite in radial direction and 
locality is only gained in longitudinal direction. To examine this, 
a $\delta$-peak in E and B is initialized on axis in the center of the 
simulation box with a grid resolution of $\Delta r = \Delta z = 1 \, \mu \mathrm{m}$. 
It is then propagated for one timestep $\Delta t = \Delta z /c$. 
Fig.\ \ref{fig:stencil_reach} (top) shows the 
expansion of the pulse for a stencil order of 32 compared to the standard 
PSATD. When using the unmodified PSATD ($\text{PSATD}_\infty$) scheme the 
simulation shows field contributions along the entire axis, whereas the 
fields modeled with finite order PSATD quickly decrease in amplitude and 
reach the machine precision level. The expansion of the unit pulse strictly 
follows the stencil coefficients $\alpha*$ derived from the modified PSATD 
operations given by
\begin{equation}
  \alpha^*=\sqrt{|\mathcal{F}^{-1}(C)^2|+|\mathcal{F}^{-1}(Sk_\perp)^2|+|\mathcal{F}^{-1}(S[k_z])^2|},
\end{equation}
where $\mathcal{F}^{-1}$ denotes an inverse Fourier transform along the $z$-axis. Here these 
coefficients are calculated numerically and instead of an analytical Fourier 
transform an FFT was used. The coefficients are evaluated for $k_\perp=0.5 \, \mu \mathrm{m}^{-1}$.
The stencil coefficients as well as the unit pulse reach machine precision 
related noise after decreasing by $16$ orders of 
magnitude over a range of $\Delta N_z$ cells. In order to show how the stencil 
extent $\Delta N_z$ scales with the arbitrary order PSATD, the same simulation 
is done with varied stencil order. Fig.\ \ref{fig:stencil_reach} (bottom) shows 
that the stencil extent behaves non-linearly to the applied stencil order. While 
for high orders the extent is similar to the stencil order, the reach of the 
solver tends to be bigger than the order in low order cases.
\\In a domain decomposition scheme the fields would only need to be exchanged 
between neighboring processes in guard regions with the size of $\Delta N_z$ 
cells, as this is the maximum distance information can travel within one 
timestep. These guard cells hold copies of the fields in the neighboring domains. 
At each timestep the fields are evaluated independently in each domain and the 
information is only exchanged in these cells.
When significantly less than $\Delta N_z$ guard cells are used, 
truncation of the stencil can lead to spurious fields \cite{Vincenti}.\\
Note that for parallelization also a localized current correction or a charge 
conserving current deposition is needed. For that an exact current deposition 
in k-space for Cartesian coordinates \cite{Vay2013260} and a local FFT based 
current correction on staggered grids \cite{li2016controlling} have been 
shown.\\
To summarize, in high order FDTD schemes the computational demand scales with the stencil order as an extension of the stencil to more neighboring cells calls for more individual 
numerical operations. In contrast, the stencil in spatial space for the 
arbitrary order PSATD is merely determined
by the modification of the $k$ values in spectral space. Apart from this the scheme is identical
to the infinite order PSATD. Therefore, the computational demand of the solver is independent of the stencil order and consequently also of the precision of the field solver.
However, the improved locality of this schemes comes along with approximations 
on the integration of the electromagnetic fields which also result in spurious numerical dispersion. The implications of this will be discussed in the following.

\section*{Numerical Cherenkov
radiation}\label{numerical-cherenkov-radiation}

\begin{figure}[tb]
\centering
\includegraphics[width=\columnwidth]{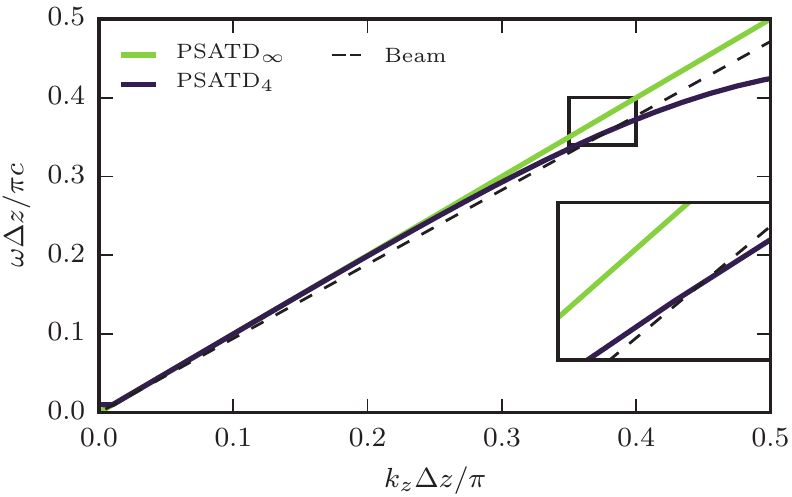}
\caption{Dispersion relation for the arbitrary order PSATD solver (solid lines) 
on the $k_z$-Axis ($k_r=0$). Finite stencils introduce a spurious dispersion 
relation, which leads to a reduced phase velocity $v_{\phi,z}=\omega/k_z$ of 
modes with large $k_z$ and allows unphysical intersections (box) with the mode 
corresponding to the velocity of a relativistic particle (dashed line), here 
$\gamma_\text{beam}=3$. Particle beams can then travel with the same velocity 
as electromagnetic modes, which leads to coherent emission of spurious radiation, 
referred to as NCR.}
\label{fig:stencil}
\end{figure}
\begin{figure}[tb]
\centering
\includegraphics[width=\columnwidth]{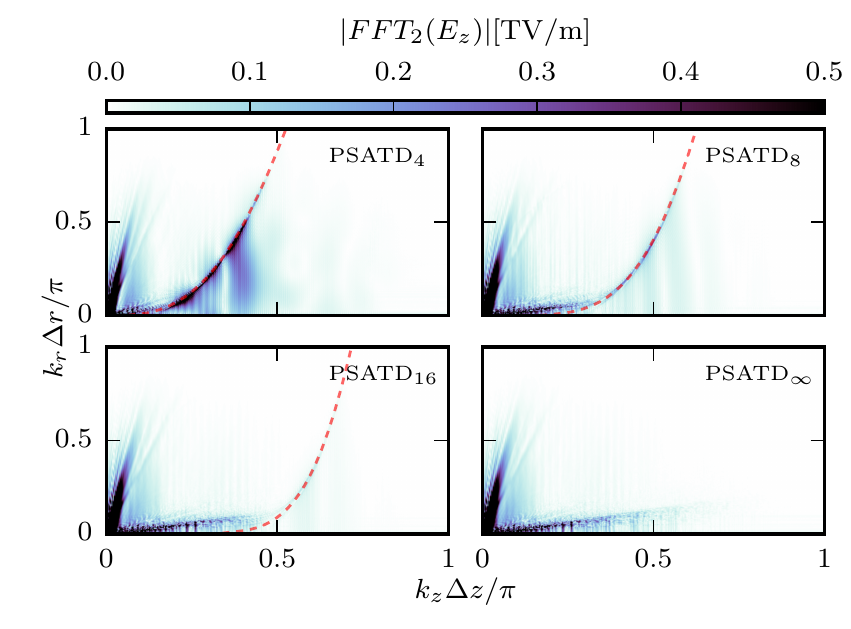}
\caption{Spatial Fourier transforms of the accelerating field for simulations 
with different PSATD stencil orders. The simulations with finite order show resonances around the modes with 
a phase velocity matching the velocity of the electron beam (dashed lines) 
given by Eq.\ \eqref{eq:resonance_k}. The magnitude of NCR decreases for 
increasing stencil order. The NCR-free infinite order PSATD simulation is shown 
for reference.}
\label{fig:FFT}
\end{figure}

In general, the phase velocity $v_\phi$ of an electromagnetic mode with wave 
vector $\boldsymbol{k}$ and angular frequency 
$\omega$ is \mbox{$v_\phi^2=\omega^2/\boldsymbol{k}^2$}.
The allowed electromagnetic modes in vacuum are described by the dispersion 
relation
\begin{equation}
  \omega^2=c^2\boldsymbol{k}^2.
  \label{eq:physical_disp}
\end{equation}
When modifying $\boldsymbol{k}$ in longitudinal direction the dispersion relation 
is
\begin{equation}
  \omega^2=c^2\left(k_r^2+[k_z]^{2}\right).
  \label{eq:omega_mod}
\end{equation}
Here, a spurious numerical dispersion is introduced, where the phase velocity 
in vacuum decreases for modes with increasing $k_z$. Fig.\ \ref{fig:stencil} 
shows the distorted dispersion relation caused by the arbitrary order PSATD 
solver compared to the accurate case (Eq.\ \eqref{eq:physical_disp}) given by 
the unmodified, infinite order PSATD method.
Due to the distortion of the dispersion relation, relativistic particles can 
travel with the same velocity as some electromagnetic modes.
This can lead to coherent emission of spurious radiation referred to as 
numerical Cherenkov radiation.
Consequently, the affected modes obey
\begin{equation}
  v_{\phi,z}=v_\text{beam},
\label{eq:resonance}
\end{equation}
where $v_\text{beam}$ is the velocity of the charged particles in the 
simulation. Using Eq.\ \eqref{eq:omega_mod} this can be rewritten as
\begin{equation}
  v_\text{beam}=\frac{c\sqrt{k_r^2+[k_z]^{2}}}{k_z}.
  \label{eq:resonance_k}
\end{equation}
This condition describes the modes, where the spurious electromagnetic 
dispersion relation intersects with the beam mode corresponding to the velocity 
of a charged particle.
This effect is illustrated in Fig.\ \ref{fig:FFT}. The dashed lines indicate 
the modes fulfilling Eq.\ \eqref{eq:resonance_k} and thus correspond to the 
intersection points of the dispersion relation and the beam mode as depicted 
in Fig.\ \ref{fig:stencil}. These intersection points denote the modes where 
NCR is excited. 
They are overlaid with the spatial Fourier transform of the longitudinal 
electric field from \textsc{Fbpic} simulations, with parameters as discussed in 
the next section.
The panels of Fig.\ \ref{fig:FFT} show results from simulations done with 
different stencil orders. They all show excited modes that can be attributed to 
the physics in the simulation of a beam driven plasma accelerator, e.g.\ the 
plasma wave or betatron radiation. However, compared to the unmodified $\mathrm{PSATD_\infty}$ 
scheme simulation, additional excited modes are visible around the modes prone 
to NCR. The excitations are especially strong for low solver orders. They
decrease for higher stencil orders as the resonant modes of Eq.\ \eqref{eq:resonance_k}
move away from modes that can be attributed to the physics of the simulation. 
Eventually, for the $\mathrm{PSATD_\infty}$ the phase velocity $v_\phi$ is always greater than 
$v_\mathrm{beam}$
(i.e.\ no dashed line) and
no NCR is present.\\
The analysis of the Fourier transformed electric field consequently can be used as
an indication for the presence and magnitude of NCR.
Next, we will show the influence of these unphysical modes on the properties of an electron 
beam.

\section*{Accurate modeling of plasma accelerators in finite order PSATD}\label{simulations}

\begin{figure}[tb]
\centering
\includegraphics[width=\columnwidth]{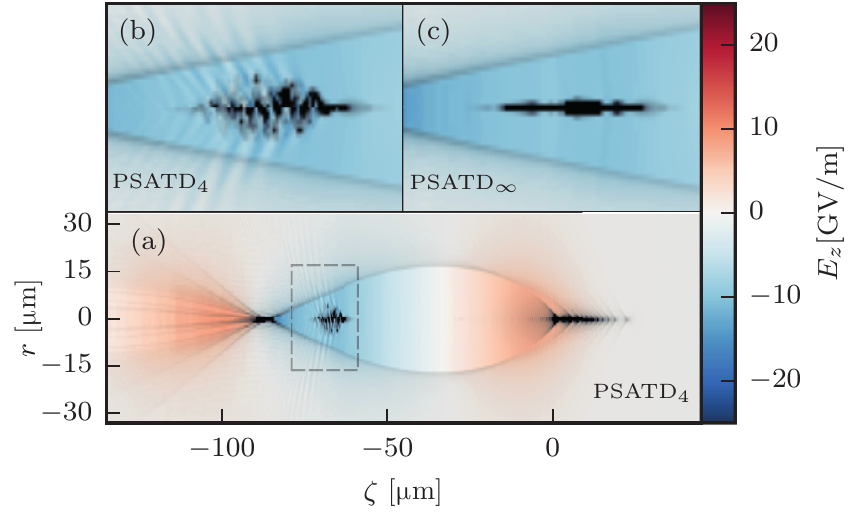}
\caption{(a) Simulation setup featuring a driver and witness beam after 
propagating $35 \, \mathrm{mm}$. Colors correspond to the accelerating field 
and the charge density is given in gray (arb. units), $\zeta = z - c t$ are 
the co-moving coordinates with $\zeta=0$ at the center of the driver beam. The 
short $\mathrm{PSATD_4}$ stencil leads to distortions of the longitudinal field 
around the witness beam (b) that are not present for an infinite order stencil 
(c).}
\label{fig:ez}
\end{figure}
\begin{figure}[tb]
\centering
\includegraphics[width=\columnwidth]{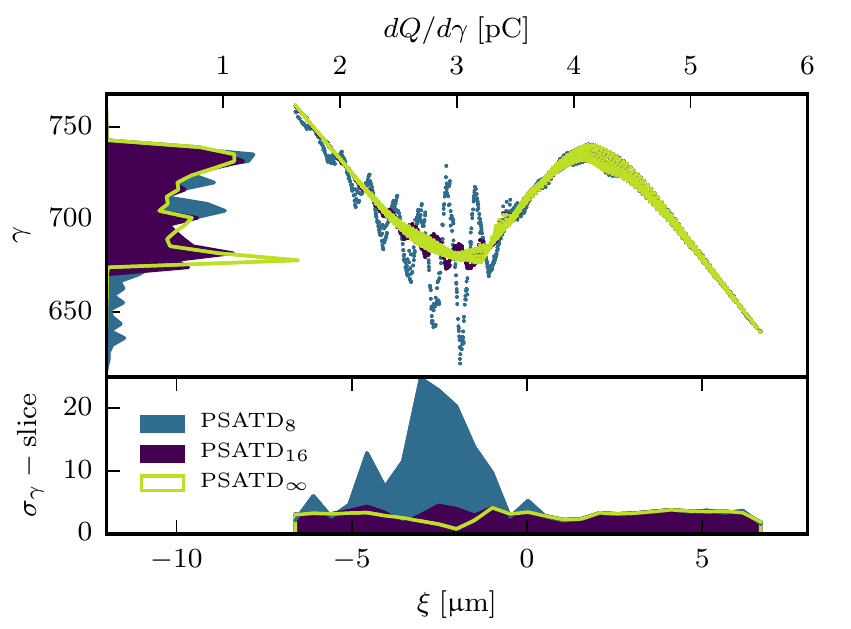}
\caption{Longitudinal phase space of the witness beam after propagation 
through $35 \, \mathrm{mm}$ of plasma. $\xi$ is the internal bunch coordinate 
with $\xi=0$ at the center of the witness beam. For decreasing stencil order 
it is increasingly spoiled by a high frequency oscillation. This causes a 
growth of projected and slice
energy spread (slice length of $0.5\; \mathrm{\mu m}$ here).}
\label{fig:psp}
\end{figure}
\begin{figure}[tb]
\centering
\includegraphics[width=\columnwidth]{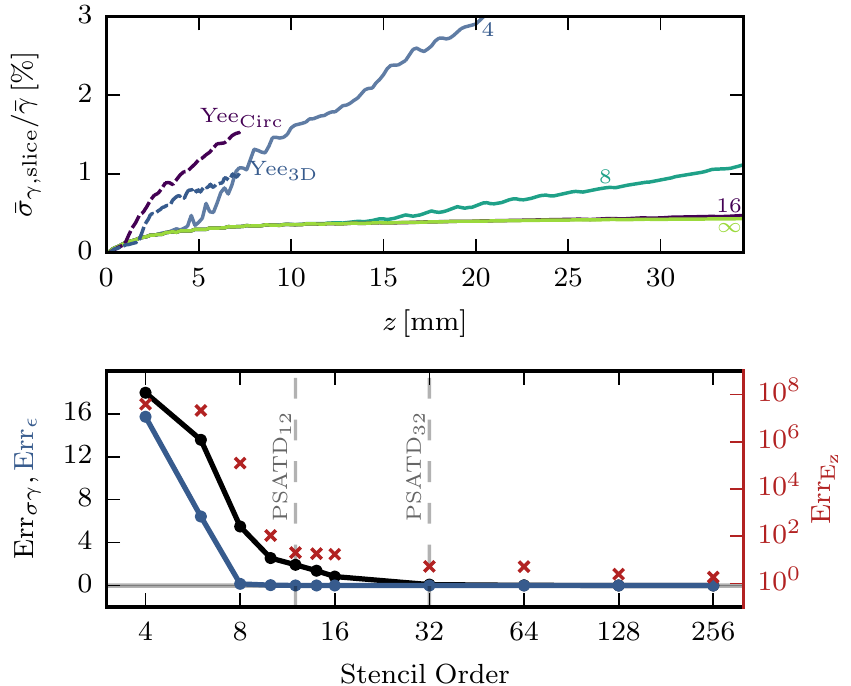}
\caption{Top: Growth of the witness beam's slice energy spread during the 
propagation through $35 \, \mathrm{mm}$ of plasma. The slice energy spread is 
averaged over all slices within $\pm$ 3 standard deviations around the bunch 
center. The solid and dashed lines show data obtained with the PSATD and Yee 
scheme, respectively. The PSATD$_{16}$ and PSATD$_\infty$ lines almost overlay 
each other. Bottom: Errors in emittance (blue), slice energy spread (black) and longitudinal field (red crosses) after $35 \, \mathrm{mm}$. For a stencil order increasing from 4 to 256 the error introduced
by NCR in the electric field decreases by 8 orders of magnitude. Consequently, the error of the final 
emittance and slice energy spread quickly approaches zero.}
\label{fig:stencil_qual}
\end{figure}
\begin{figure}[tb]
\centering
\includegraphics[width=\columnwidth]{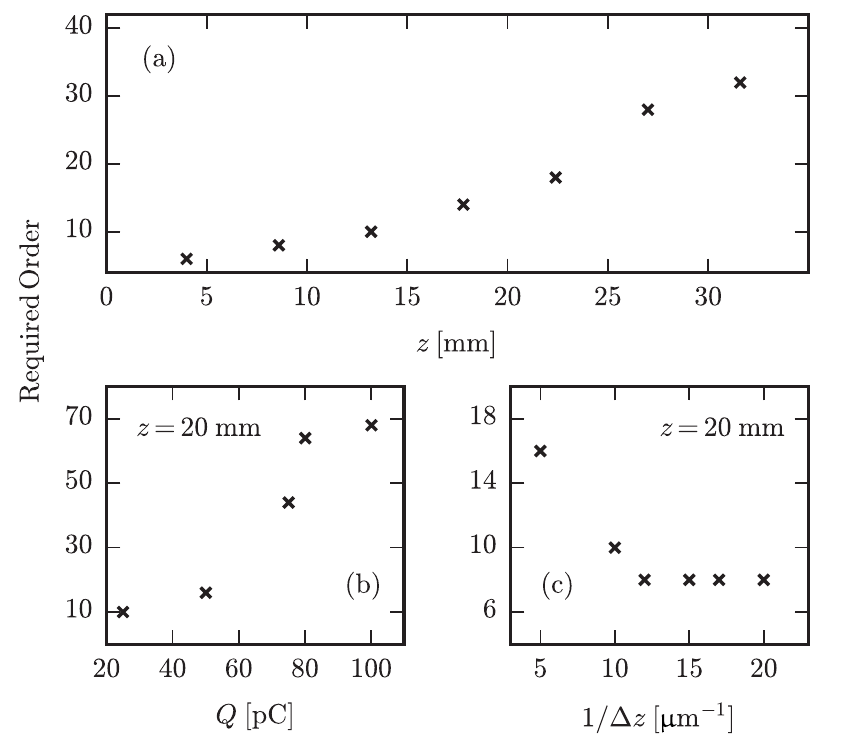}
\caption{Starting from the parameters used in Fig.\ \ref{fig:FFT}-\ref{fig:stencil_qual}
either the charge, the propagation length or the longitudinal grid resolution
is scanned and the required stencil order to fulfill $\mathrm{Err}_{\sigma\gamma}<0.05$
is shown. Top: Required stencil order depending on the propagation length for a beam charge
of $50$ pC and a resolution of $5$ cells/$\mu$m.
Bottom: Required stencil order depending on the beam charge (b) with a resolution of $5$ cells/$\mu$m, and depending on the longitudinal grid resolution (c) with a beam charge of $50$ pC, both 
evaluated after $20$ mm of propagation in plasma. }
\label{fig:parameter_scan}
\end{figure}

To show that already low stencil orders are sufficient to mitigate errors due 
to NCR in the context of plasma-based acceleration, PIC simulations with 
the spectral, quasi-cylindrical code \textsc{Fbpic} for  
beam-driven plasma accelerator parameters are presented.
Typical setups \cite{Grebenyuk2014246,Litos} feature high bunch 
charges and consequently 
high currents, as they aim for applications in future brilliant light sources 
and high energy physics.  \\
The propagation of cold, i.e.\ zero emittance, monoenergetic driver and witness 
electron beams through a plasma of density 
$1 \times 10^{17} \; \mathrm{cm^{-3}}$ is simulated. A non-linear wakefield is 
driven by a \mbox{1 GeV} Gaussian beam with a charge of \mbox{180 pC}, 
$8.4 \; \mathrm{\mu m}$ rms length and $5 \; \mathrm{\mu m}$ rms width, 
compare Fig.\ \ref{fig:ez}(a). The \mbox{25 MeV} Gaussian witness bunch of 
\mbox{50 pC} charge, $2 \; \mathrm{\mu m}$ rms length and $1 \; \mathrm{\mu m}$ 
rms width is positioned in the accelerating region at the back of the bubble.\\
The simulation box has a resolution of \mbox{5 cells/$\mathrm{\mu m}$}
longitudinally and \mbox{1.5 cells/$\mathrm{\mu m}$} transversally, with 16 
particles
per cell distributed as $2\times  2  \times  4$ in longitudinal, radial and 
azimuthal direction, respectively. A total of 900 $\times$ 300 cells in two 
azimuthal modes is used.
While the pseudo-spectral solver of \textsc{Fbpic} is technically not limited by a CFL 
condition, a timestep of $\Delta t = \Delta z /c$ was used. 
For the same physical parameters, the stencil order 
of the arbitrary order PSATD is varied.
The NCR-free simulation with an infinite order stencil is used as a reference 
in the following.\\
Fig.\ \ref{fig:ez} and \ref{fig:psp} both show simulation results after a 
propagation through \mbox{35 mm} of plasma. In Fig.\ \ref{fig:ez} (b) and (c) 
the accelerating field $E_z$ and the charge density $\rho$ in the vicinity of 
the witness bunch can be seen. In the case of a low stencil order strong 
distortions are visible in $E_z$. The distortions are the spatial space equivalent
to resonances visible for the $\mathrm{PSATD_4}$ in Fig.\ \ref{fig:FFT}.
These spurious fields affect the witness beam shape which
corresponds to an increase of beam emittance, as it has been reported for 3D 
FDTD codes \cite{lehe2013}.\\
We find that also the longitudinal phase space is affected, as shown in Fig.\ 
\ref{fig:psp}, which compares the phase spaces modeled with several stencil 
orders.
The projected energy spread increases during propagation through the plasma due 
to the slope of $E_z$ in combination with beam loading, so that an s-shape like 
structure is imprinted on the phase space also with $\mathrm{PSATD_\infty}$.
However, for decreasing stencil order, NCR causes an increasingly violent high 
frequency oscillation on the longitudinal phase space, leading to an unphysical growth of 
both slice and projected energy spread. It can be observed that the 
perturbation is especially pronounced behind the beam center, where the current 
density is highest, while the head remains mostly unaffected.\\
Fig.\ \ref{fig:stencil_qual} (top) shows the evolution of the slice energy 
spread during the propagation through the plasma for different solver orders. 
For finite order stencils the energy spread grows non-linearly with $z$. 
While $\mathrm{PSATD_4}$ and $\mathrm{PSATD_8}$ still deviate significantly 
from the Cherenkov free infinite order case, the impact of NCR on the beam in 
the PSATD$_{16}$ simulation is barely visible even after a long propagation 
distance. The dashed lines correspond to simulations done with the FDTD Yee 
scheme in 3D Cartesian and quasi-cylindrical geometry. These are performed with 
the PIC code \textsc{Warp} \cite{warp} using the same parameters and resolution 
as \textsc{Fbpic}. With the Yee scheme the growth rate of spurious slice energy 
spread is even greater than for the low order PSATD case. Note that the 
\textsc{Warp} framework also features a $\mathrm{PSATD_\infty}$ (and arbitrary order) solver, which, 
if used for this problem, would also produce the correct physical results as the 
\textsc{Fbpic} $\mathrm{PSATD_\infty}$ algorithm.\\
Fig.\ \ref{fig:stencil_qual} (bottom) shows the error of the slice energy spread and the 
transverse beam emittance and the longitudinal electric field after a propagation 
length of \mbox{35 mm} depending 
on the stencil order. Here, the errors are defined as

\begin{align}
  &\mathrm{Err}_{\sigma\gamma}=\sqrt{\frac{1}{N_\mathrm{slice}}\sum_i^{N_\mathrm{slice}}\left(\frac{\sigma_{\gamma,f,i}-\sigma_{\gamma,\infty,i}}{\sigma_{\gamma,\infty,i}}\right)^2},\\
  &\mathrm{Err}_\epsilon=\frac{\epsilon_f-\epsilon_\infty}{\epsilon_\infty},\\
  &\mathrm{Err}_{E_z}=\sqrt{\frac{1}{N_zN_r}\sum_{N_z,N_r}\left(\frac{FFT_2(E_z)_f-FFT_2(E_z)_\infty}{FFT_2(E_z)_\infty}\right)^2},
\end{align}

respectively. The subscripts $f$ and $\infty$ indicate quantities from finite and 
infinite order simulations. For the calculation of $\mathrm{Err}_{\sigma\gamma}$ $N_\mathrm{slice}$ slices
of length $0.5$ $\mu$m within $\pm3\sigma_z$ of the witness beam are considered.
For $\mathrm{Err}_{E_z}$ the sum is calculated over all cells of the spatial Fourier transform
of $E_z$ (compare Fig.\ \ref{fig:FFT}). This quantity can be regarded as measure
for the strength of the spurious fields caused by NCR.\\
As can be expected from Fig.\ \ref{fig:FFT}, Fig.\ \ref{fig:stencil_qual} (bottom)
confirms that the unphysical contributions of NCR to the field, $\mathrm{Err}_{E_z}$, 
depend strongly on the stencil order and decrease by 8 orders of magnitude as 
the stencil order increases from 4 to 256. With the strength of NCR decreasing,
the beam quality in terms of energy spread and emittance also converges toward the 
correct values given by the reference case with a PSATD$_\infty$.
The beam's energy spread is even more sensitive to 
potential NCR than the beam emittance. $\mathrm{Err}_\epsilon$
and $\mathrm{Err}_{\sigma\gamma}$ are below 0.05 for the stencil orders of 12 and 32, respectively. In the following, this will also be used as a criterion for sufficient accuracy.\\
The stencil order necessary to meet this criterion depends on the physical and numerical parameters
of the simulation. Fig.\ \ref{fig:parameter_scan} is obtained from scans over the beam charge, the propagation length as well as the longitudinal grid resolution and shows the
required stencil order to achieve $\mathrm{Err}_{\sigma\gamma}<0.05$. 
Fig.\ \ref{fig:parameter_scan} (top) shows the required stencil order depending
on the propagation length in plasma for the same parameters
used in Fig.\ \ref{fig:FFT}-\ref{fig:stencil_qual}. For a reduced propagation 
length accurate modeling is possible with lower stencil orders. 
This is a result of the difference in the growth rate of
the spurious slice energy spread seen in Fig.\ \ref{fig:stencil_qual} (top).\\
NCR also scales with the bunch charge density and the longitudinal grid resolution.
The connection of the required stencil order with these parameters is shown in Fig.\ \ref{fig:parameter_scan} (bottom). As the computational demand increases quadratically with
the grid resolution, this scan and for consistency also the charge scan are evaluated after a 
propagation length of $20$ mm.
Clearly, high charges call for increased stencil orders (see Fig.\ \ref{fig:parameter_scan}b). However, please note
that in the presented case, due to the small beam dimensions $50$ pC already 
correspond to a high peak current of $3$ kA and a peak electron density 
of $1 \times 10^{19}\; \mathrm{cm}^{-3}$.\\
The required stencil order can be reduced by increasing the longitudinal grid resolution
(see Fig.\ \ref{fig:parameter_scan}c).
However, for the propagation length and charge considered here, it is necessary to apply 
a stencil order of at least 8 even when using a resolution of 20 cells/$\mu$m.

\section*{Conclusion}\label{conclusion}

In conclusion, NCR stemming from erroneous modeling of the dispersion relation 
leads to beam phase space degradation, which manifests itself in a spurious 
growth of slice energy spread and emittance. This effect is present in both the 
FDTD Yee scheme and finite order PSATD solvers. Errors from NCR rapidly 
decrease for higher stencil orders, eventually showing the same results as the 
NCR-free infinite order PSATD solver.\\
In the case discussed here, which is inspired by typical beam driven plasma 
acceleration parameters, a stencil order of 32 effectively suppresses spurious 
beam quality degradation. 
We have chosen a long propagation distance of \mbox{35 mm} and a high witness beam charge of \mbox{50 pC} corresponding to an electron density of $1 \times 10^{19}$ cm$^{-3}$.
This is a conservative example in the sense that our results are applicable to 
many setups in the field of plasma acceleration that use more moderate parameters. 
This is confirmed by parameter scans that show that the constraints on the required stencil order 
are relaxed for a lower beam charge and shorter propagation length.\\
In practice, this means that with around $60$ guard cells between neighboring 
domains, parallelization is possible while retaining the physics of the 
considered problem. \\
Analogous to the frequently used Yee scheme, growth of NCR in the finite order 
stencil PSATD scheme can also be reduced by increasing the resolution of the 
simulation grid. However, this is an inefficient solution especially in schemes 
where the timestep is limited by a CFL-condition and hence directly linked to 
the grid resolution. Yet, this means that in simulations with high spatial 
resolution the effects of NCR are less severe. This, for example, would be the 
case for laser plasma acceleration, where a high resolution is required in any case
to resolve the laser wavelength. Therefore, in these cases even lower order 
stencils than suggested here are sufficient to model the physics without 
artifacts from NCR. \\
The arbitrary order PSATD scheme preserves the benefits of pseudo-spectral 
solvers, e.g.\ the analytic integration in time or a centered definition of the 
electric and magnetic fields. It further allows to independently adjust the precision of the 
electromagnetic dispersion relation and the resolution of the simulation. 
This way the needed spatial resolution is governed by the physical problem 
and not by the mitigation of NCR.\\

Input scripts to reproduce the presented \textsc{Fbpic} and \textsc{Warp} simulations
can be found in Ref.\ \cite{scripts}.

\begin{acknowledgments}
We gratefully acknowledge the computing time provided on the supercomputer 
JURECA under project HHH20. Work at LBNL was funded by the Director, 
Office of Science, Office of High Energy Physics, U.S. Department of 
Energy under Contract No. DE-AC02- 05CH11231, including the Laboratory Directed 
Research and Development (LDRD) funding from Berkeley Lab.
\end{acknowledgments}

\bibliography{cites}
\end{document}